\documentclass[prl,twocolumn,showpacs,amsmath,amssymb]{revtex4}
\usepackage{graphics}
\usepackage{amsmath}

\usepackage[usenames,dvipsnames]{color}

\def\tmo{TbMnO$_3$}

\begin{document}
\title{First-principles theory of magnetically induced ferroelectricity in \tmo}
%\subtitle{Do you have a subtitle?\\ If so, write it here}

\author{Andrei Malashevich}
\email{andreim@physics.rutgers.edu}
%% CUT?
%
\author{David Vanderbilt}
\affiliation{
Department of Physics \& Astronomy, Rutgers University,
Piscataway, NJ 08854-8019, USA
}

\date{\today}

\begin{abstract}
%Insert your abstract here.
We study the polarization induced via spin-orbit interaction
by a magnetic cycloidal order
in orthorhombic \tmo\ using first-principle methods. The case
of magnetic spiral lying in the $b$-$c$ plane is analyzed,
in which the pure electronic contribution to the polarization
is shown to be small. We focus our attention on the lattice-mediated
contribution, and study it's dependence on the Coulomb interaction
parameter $U$ in the LDA+$U$ method and on the wave-vector of the 
spin spiral. The role of the spin-orbit interaction on different sites
is also analyzed.
\end{abstract}

% PACS, the Physics and Astronomy
                             % Classification Scheme.
\maketitle
\section{Introduction}
\label{intro}

Magnetically induced ferroelectricity provides a route to materials
with a large magneto-electric (ME) coupling. The appearance of the
ferroelectric order simultaneously with the onset of a particular
magnetic order suggests that the magnetic and electrical properties
should be strongly interconnected. Therefore, one might expect to observe
a strong dependence of the electric polarization on external magnetic
fields in such materials. Indeed, a pronounced interplay between
ferroelectricity and magnetism was observed experimentally
\cite{kimura,lottermoser,hur,lawes:2005}
in (Gd,Tb,Dy)MnO$_3$, (Tb,Dy)Mn$_2$O$_5$, Ni$_3$V$_2$O$_8$, etc.
In these materials, ferroelectricity is induced by a specific (e.g.,
cycloidal) magnetic ordering.
Since polarization is a polar vector quantity, it is necessary to break
the inversion symmetry in order for the ferroelectric phase to appear.
In magnetically induced ferroelectrics, the inversion symmetry can be
broken by the magnetic order alone, or in some cases, by the magnetic
order in combination with the crystal structure (e.g., in
Ca$_3$CoMnO$_6$~\cite{cheong:2007}), where separately both the
lattice and magnetic structures have inversion symmetry.
The spin-orbit (SO) coupling is essential in order to
allow the broken symmetry to be communicated from the magnetic
to the charge and lattice degrees of freedom.

In our work, we focus our attention on \tmo, in which the appearance 
of polarization is attributed to the cycloidal order of magnetic moments
on the Mn$^{3+}$ sites.
\tmo\ is an orthorhombically distorted perovskite material whose
lattice has {\it Pbnm} symmetry with 20 atoms per unit cell.
The magnetic properties of \tmo\ are mainly due to the Mn$^{3+}$ ions.
The competition between the nearest-neighbor and next-nearest-neighbor
spin interactions leads to a rich phase diagram. Experimentally it is
known that below $\sim{41}$~K the magnetic moments form 
a sinusoidally modulated antiferromagnetic order. Below
$\sim{27}$~K, a phase transition occurs in which the magnetic order
develops a cycloidal character (with spins in the $b$-$c$ plane)
with modulation wave-vector $k_s\sim0.28$ along $b$, and a
polarization simultaneously appears along $c$.
Phenomenological and microscopic theoretical
models~\cite{most:2006,katsura:2005,serg:2006}
can explain how the cycloidal magnetic order can drive the system
to become ferroelectric, and one can predict the direction of the
polarization (but not its sign) on rather general symmetry grounds.
However, it is unclear whether one should expect the ionic
displacements (phonons) to play the most significant role in creating
electric polarization, or whether a purely electronic mechanism could
explain the observations.

Our previously-reported initial findings~\cite{malash} and a related
study~\cite{Xiang}
suggested that, at least in the case of the spin spiral lying in the
$b$-$c$ plane, the lattice contribution to the polarization dominates
over the electronic contribution in \tmo.  However, these works left
several questions unanswered.  Here, we first briefly review our earlier
work, emphasizing a careful analysis of the lattice-mediated contribution
to the polarization using first-prin\-ci\-ples calculations.
We then discuss several extensions of the work,
including a study of the dependence of the results on the choice of
$U$ parameter, additional details concerning the site-specific spin-orbit
interaction and its effects on the dynamical effective charges, and
a more thorough and revealing investigation of the dependence of the
polarization on the wave-vector of the spin spiral.

\section{Computational details}
\label{comp}

\begin{figure}
\resizebox{0.5\textwidth}{!}{%
\includegraphics{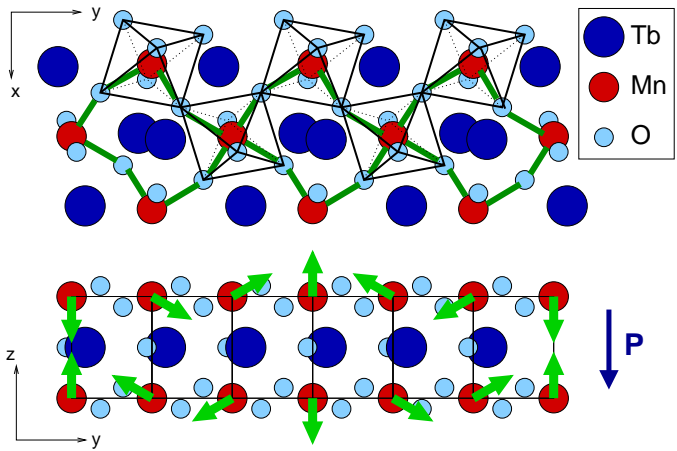}
}
\caption{Sketch of the $a\times3b\times c/2$ orthorhombic cell
of \tmo\ ({\it Pbnm}
space group) showing MnO$_6$ octahedral tilts (top)
and the cycloidal magnetic structure on the Mn$^{3+}$ sites (bottom).}
\label{model}       % Give a unique label
\end{figure}

The electronic-structure calculations are based on the density functional
theory (DFT) and are performed using the
projector-augmented wave (PAW) method~\cite{Blochl,Kresse:1999} 
in a plane-wave basis set. The plane-wave energy cutoff is 500 eV.
We use the VASP code package~\cite{kresse-vasp}
for our calculations. We do not include $f$ electrons in the Tb PAW potential.
The local-density approximation for the exchange-correlation
functional is used (Ceperley-Alder~\cite{Ceperley} 
with Vosko-Wilk-Nusair correlation~\cite{voskown}). For a proper treatment
of Mn $d$ electrons, we use on-site Coulomb corrections implemented
in a rotationally-invariant LDA+$U$ formulation~\cite{Dudarev}.
We consider two values of $U$, 1 and 4 eV.
The results of a systematic study with $U=1$\,eV were reported previously
in Ref.~\cite{malash}; some of these will be reproduced here for 
comparison. 
We consider the cycloidal magnetic order having a spin spiral
lying in the $b$-$c$ plane and wave-vector $k_{\rm s}$ propagating
in the $b$-direction (see Fig.~\ref{model}). The experimental value of
the wave vector is incommensurate with the lattice, 
$k_{\rm s}\sim0.28$~\cite{kimura}. Since we use periodic boundary conditions,
we study commensurate cycloidal spin structures with wave-vectors
$k_{\rm s}=0,1/4,1/3,1/2,2/3$ and 1 using appropriate supercells. 
The structural relaxation was performed with $k_{\rm s}=1/3$ (60 atoms per cell),
which is close to the experimental $k_{\rm s}$ (Fig.~\ref{model}).
A 3$\times$1$\times$2 $k$-point sampling is used.
The Berry-phase approach~\cite{King-Smith} is used for the calculation of the 
electric polarization.

\section{Results}
\label{results}

\subsection{Dependence on $U$ parameter}
\label{Udep}

A crystal structure optimization was performed for the 60-atom supercell
using $U=1$\,eV and $U=4$\,eV. These calculations were carried out without
SO interaction to obtain centrosymmetric reference structures, with respect
to which Berry phase polarization will be calculated in the subsequent analysis.
The lattice parameters and Wyckoff coordinates of the relaxed structures
are given in Table~\ref{tab:str}.
For both values of $U$, the calculated structural parameters are very close
to the experimental ones, with the lattice vectors being in slightly better
agreement with experiment for the case of $U=4$\,eV.

The calculation of the polarization
in the presence of the SO interaction for these relaxed structures
yields an estimate for the pure electronic contribution to the polarization.
We find $P^{\rm elec}=32\,\mu{\rm C}/{\rm m}^2$ and
$P^{\rm elec}=-14\,\mu{\rm C}/{\rm m}^2$ for $U=1$\,eV and $U=4$\,eV
respectively (with the direction of polarization parallel to the $c$ axis,
as was expected from symmetry). 
In both cases, the magnitude is much smaller than the observed
value of $\sim600\,\mu{\rm C}/{\rm m}^2$~\cite{kimura}, indicating that it can
be neglected for any reasonable value of $U$.

\begin{table}
\caption{Experimental (Ref.~\cite{alonso})
and theoretical ($U$=1\,eV is taken
from Ref.~\cite{malash}) 
structural parameters for orthorhombic \tmo.}
\label{tab:str}       % Give a unique label
\begin{tabular}{clcccc}
\hline\noalign{\smallskip}
 & & & Exp. & $U=1$\,eV & $U=4$\,eV \\
\noalign{\smallskip}\hline\noalign{\smallskip}
\multicolumn{2}{l}{Lattice vectors}
   & $a$ (\AA)                 & 5.293 & 5.195 & 5.228 \\
 & & $b$ (\AA)                 & 5.838 & 5.758 & 5.775 \\
 & & $c$ (\AA)                 & 7.403 & 7.308 & 7.343 \\
Tb & 4c($x$ $y$ 1/4) & $x$     & 0.983 & 0.979 & 0.980 \\
   &                 & $y$     & 0.082 & 0.084 & 0.084 \\
Mn & 4b(1/2 0 0)     & & & \\
O1 & 4c($x$ $y$ 1/4) & $x$     & 0.104 & 0.107 & 0.111 \\
   &                 & $y$     & 0.467 & 0.469 & 0.465 \\
O2 & 8d($x$ $y$ $z$) & $x$     & 0.704 & 0.699 & 0.700 \\
   &                 & $y$     & 0.326 & 0.320 & 0.323 \\
   &                 & $z$     & 0.051 & 0.052 & 0.053 \\
\noalign{\smallskip}\hline
\end{tabular}
\end{table}

To analyze thoroughly the role of the ionic displacements,
we computed the Hellmann-Feynman forces which appeared on the ions
after the SO interaction was turned on.
Keeping in mind that only zone-center infra-red active
modes can be responsible for the appearance of polarization,
we filtered out the forces that did not act on these modes.
We found that the infra-red active modes that were left
had dynamical dipoles along $c$ as expected.
There are eight such modes:
three associated with Mn atoms in Wyckoff position 4b,
one each for Tb and O1 atoms in Wyckoff position 4c,
and three for the O2 atoms in Wyckoff position 8d \cite{bcs:sam}.
To find the ionic displacements induced by the SO interaction,
we used the force-constant matrix~\cite{malash}
calculated in the absence of SO using the 60-atom
supercell. (We note however that one could have used the 20-atom
unit cell for this calculation, with $k_{\rm s}=0$, since
the force-constant matrix depends only weakly on the wave-vector.)
Updating the ionic positions according to the calculated
displacements and carrying out Berry-phase polarization calculations
for the resulting structures, we find the total
polarization induced by the SO interaction. The results for
$U=1$\,eV and $U=4$\,eV are 
$P^{\rm tot}=-467\,\mu{\rm C}/{\rm m}^2$ and
$P^{\rm tot}=-218\,\mu{\rm C}/{\rm m}^2$ respectively.
Knowing the ionic displacements, one can also find the 
contributions to the polarization from each mode by calculating
the effective charges of the modes. The results of such calculations
for both values of $U$ are given together with the forces in
Table~\ref{tab:wyck_modes}.

Note that the values for the forces in the second column of
the table represent the forces acting on the `modes'
rather than ions themselves. Each mode is characterized by 
a unit vector in the $(20\times3)$-dimensional space.
Therefore, to get the actual number for the force acting on a 
particular ion, one must multiply the corresponding value
from the table by an appropriate unit vector. This procedure
will give absolute values for the ionic forces that are two times
smaller than the numbers in the table for the first five modes,
and $2\sqrt{2}$ times smaller for the rest of the modes.

\begin{table}
\caption{Forces $F$ (meV/\AA), effective charges $Z$ ($e$),
and contributions $\Delta P$ to the polarization ($\mu$C/m$^2$)
from IR-active modes.
See text for the description of the conventions used to describe the modes.
The values for $\Delta{P}^{*}$ are calculated with the SO coupling turned
off everywhere except on Mn sites.}

\label{tab:wyck_modes} 
\begin{tabular}{lrrrr|rr}
\hline\noalign{\smallskip}
Wyckoff & \multicolumn{4}{c}{$U=1$\,eV} & \multicolumn{2}{c}{$U=4$\,eV} \\  
~~mode  & $F$~~~ & \!$Z$ & $\Delta{P}$~~ & $\Delta{P}^{*}$~ & $F$~~~ & \!\!$\Delta{P}$ \\ 
\noalign{\smallskip}\hline\noalign{\smallskip}
 Tb 4c, $z$ & \!\! 0.43   & \!   7.47 & \!\!$-$94  & \!\!$-$73  &    0.47 & \!\!$-$42 \\
 O1 4c, $z$ & \!\! 2.26   & \!$-$6.82 & \!\!   81  & \!\!   69  &    1.46 & \!\!   15 \\
 Mn 4b, $x$ & \!\!$-$7.04 & \!   0.57 & \!\!$-$13  & \!\!$-$14  & $-$2.00 & \!\!$-$3  \\
 Mn 4b, $z$ & \!\!$-$8.93 & \!   7.46 & \!\!$-$248 & \!\!$-$232 & $-$3.86 & \!\!$-$94 \\
 Mn 4b, $y$ & \!\!$-$2.94 & \!   0.55 & \!\!   0   & \!\!$-$1   & $-$1.06 & \!\!   2  \\
 O2 8d, $x$ & \!\! 5.06   & \!   0.08 & \!\!    3  & \!\!   2   &    2.00 & \!\!   2  \\
 O2 8d, $y$ & \!\! 3.57   & \!   0.23 & \!\!   16  & \!\!   9   &    1.59 & \!\!   6  \\
 O2 8d, $z$ & \!\! 4.41   & \!$-$5.74 & \!\!$-$234 & \!\!$-$208 &    1.37 & \!\!$-$87 \\
\noalign{\smallskip}\hline\noalign{\smallskip}
 & & & \!\!$-$489 & \!\!$-$448 & & \!\!$-$201 \\
\noalign{\smallskip}\hline
\end{tabular}
\end{table}

The decomposition of the lattice-mediated polarization into mode contributions
is discussed in detail in Ref. \cite{malash}. Here we compare the results
obtained with different Coulomb interaction parameters.
The results in Table~\ref{tab:wyck_modes}
for $U=1$\,eV and $U=4$\,eV may seem different, 
but in fact they are qualitatively similar. The forces may be viewed as vectors
in an 8-dimensional space, and one can calculate the angle $\theta$
between them
to find $\cos\theta\sim0.98$.  Thus, the direction of the forces is almost
the same, although the magnitude is different.  This means
that the underlying physical mechanism of the magnetically induced polarization
does not depend strongly on the choice of $U$, while the magnitude of the
effect does change with $U$.  

In the subsequent sections
we will consider only $U=1$\,eV, which gives better agreement
with the experimental polarization.
Moreover, in this case the calculated band gap
matches the experimental value of $\sim0.5$\,eV~\cite{Cui:2005}.
In general, a better strategy for choosing the $U$ value would be to calculate
the exchange parameters and fit those, rather than the band gap, to the 
experimental data. In the present case, however, such an approach 
leads to a similar choice~\cite{Xiang} of parameters.

\subsection{Site-specific spin-orbit interactions}
\label{SO}

To better understand the role of the spin-orbit interaction,
we studied how the behavior of the system changes depending
on the strength of the SO interaction and on the presence of SO on 
various atomic sites. We performed
calculations of the forces with the SO interaction turned
off on all sites other than Tb, then Mn, then
O. Using the force-constant matrix, we estimated 
the lattice contributions to the polarization and found them to be
$-11\,\mu$C/m$^2$, $-447\,\mu$C/m$^2$ and $-8\,\mu$C/m$^2$,
respectively. This result shows that the SO interaction
on the Mn sites is responsible for almost all of the lattice-mediated 
contribution to the polarization (see the column for $\Delta{P}^{*}$
in the Table~\ref{tab:wyck_modes}). 
We also calculated the purely electronic contributions to the polarizations
for these three cases. Interestingly, the electronic
contribution comes almost entirely from the spin-orbit
effect on the Tb sites. Several calculations of the forces
with modified SO strength confirmed that they depend linearly on
the SO strength.

\subsection{Dependence on wave-vector}
\label{kdep}

\begin{figure}
%\resizebox{0.45\textwidth}{!}{%
\includegraphics{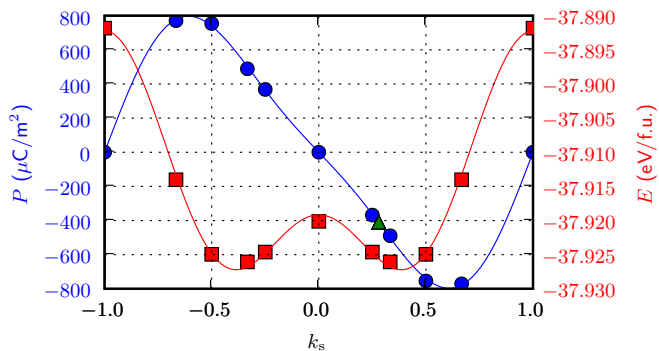}
%}
\caption{Dependence of polarization (circles, scale at left) and
total energy (squares, scale at right)
on the spiral wave-vector $k_{\rm s}$. The triangle indicates the extrapolation 
of the polarization to the experimental wave-vector of $k_{\rm s}=0.28$.}
\label{ksplot}       % Give a unique label
\end{figure}

Microscopic theoretical models 
involving the displacements of ions~\cite{serg:2006,harris,hu:2007}
show that the Dzyaloshinskii-Moriya (DM) interaction
can induce ferroelectric displacements of ions.
Usually, only the interaction between the nearest-neighbor transition
metals is considered in such models. As a consequence, the polarization
is expected to depend sinusoidally 
on the angle between the spins of the nearest-neighbor Mn sites,
${\bf P} \propto {\bf e}_{n,n'}\times({\bf S}_n\times{\bf S}_{n'})$
\cite{Kimura:2007}, and thus sinusoidally on $k_{\rm s}$.
To study the dependence of the lattice contribution to the electric 
polarization on $k_s$, we have carried calculations of the SO-induced
forces for a 40-atom supercell ($k_{\rm s}$=$1/2$)
and an 80-atom supercell ($k_{\rm s}$=$1/4$). Note also that
the same 60-atom supercell can be used to set up a spiral with 
the wave-vector $k_{\rm s}$=2/3. In general, if the supercell
consists of $n$ primitive cells, one can construct spirals
with wave-vectors $m/n$, where $0\le m\le n$. We also used the 
primitive 20-atom cell for the calculations with $k_{\rm s}=0$ and
$k_{\rm s}=1$. In all these calculations we kept the same structural 
coordinates as for the 60-atom structure to make sure that we only
change one parameter ($k_{\rm s}$) in this study.
We calculated the forces, and again filtered those that were IR-active.
We find that the pattern of the IR forces matches almost
exactly that of the 60-atom cell, i.e., the directions
of the eight-dimensional vectors almost coincide in all cases.
Using the force-constant matrix calculated on the 60-atom supercell
and the effective charges from Table~\ref{tab:wyck_modes},
we can estimate the polarization for the new wave-vectors.
The result is shown in Fig.~\ref{ksplot}. Actual calculations were
performed only for non-negative values of $k_{\rm s}$, as
symmetry arguments show the that polarization must be an odd
function of spin-spiral wave-vector. The values $k_{\rm s}=0$ and
$k_{\rm s}=1$ correspond to a collinear spin arrangement, in which
cases the polarization was zero as expected.
One can see that the dependence of the polarization on $k_{\rm s}$
deviates significantly from a simple sinusoidal form. Surprisingly,
the polarization is almost linear in $k_{\rm s}$ up to $k_{\rm s}=1/2$;
such a linear dependence is expected in the long-wavelength limit, but
that is a rather stretched assumption for $k_{\rm s}$ up to 1/2.
This result indicates that nearest-neighbor DM models oversimplify the 
mechanism of the polarization induction, and that taking the
next-nearest-neighbor interactions into account may be important.

The experimental $k_{\rm s}$ lies in the range where we can assume a linear 
dependence. The extrapolation to $k_{\rm s}$=0.28 yields a value for
the lattice contribution of the polarization of about $-410\,\mu$C/m$^2$. 
Fig.~\ref{model} also shows the dependence of the total energy per formula unit
on the wave-vector. We remind the reader
that in these calculations the structural
parameters were kept fixed, with only the directions of the magnetic moments
on Mn sites being changed. However, one can still see that the wave-vector 
at which the minimum occurs is close to the experimental value of $k_{\rm s}$.

\section{Summary}
\label{summary}

We have analyzed the lattice contribution to the electric polarization
in the cycloidal-spin compound \tmo, with the spin spiral lying 
in the $b$-$c$ plane, using density functional theory within the LDA+$U$
framework. We compare the results for two values of $U$ (1 and 4 eV),
and find that the mechanism of magnetoelectric coupling is quite 
independent of $U$. 
The dependence of the polarization on the spin-spiral wave period
is studied in detail. We find that it deviates significantly
from the sinusoidal dependence expected from simple models.
The polarization is almost linear in wave-vector for absolute
values of $k_{\rm s}$ up to 0.5.

This work was supported by NSF Grant No.~DMR-0549198.

%
% BibTeX users please use
%\bibliographystyle{}
%\bibliography{}
%
% Non-BibTeX users please use

\end{document}